\documentclass[twocolumn,aps,prb,groupedaddress]{revtex4}
\usepackage{amsmath}
\usepackage{amssymb}
\usepackage{etoolbox}
\usepackage{graphicx}
\usepackage{color}

\graphicspath{{images/}}

\newcommand{\be}{\begin{equation}}
\newcommand{\ee}{\end{equation}}
\def\ba{\begin{aligned}}
\def\ea{\end{aligned}}

\newcommand{\bea}{\begin{eqnarray}}
\newcommand{\eea}{\end{eqnarray}}

\renewcommand{\Re}{{\rm \, Re\,}}

\begin{document}

\title{Support set of random wave-functions on the Bethe lattice}
\author{A. De Luca $^1$, B. L. Altshuler $^2$, V. E. Kravtsov $^{3,4}$ and A. Scardicchio$^{3,5}$}
\affiliation{ $^1$ Laboratoire de Physique Th\'eorique de l'ENS \& Institut de Physique
Theorique Philippe Meyer\\
 24, rue Lhomond 75005 Paris - France.}
\affiliation{$^2$ Physics Department, Columbia University, 538 West 120th Street, New York, N.Y. 10027, USA}
\affiliation{ $^3$ Abdus Salam International Center
for Theoretical Physics, Strada Costiera 11, 34151 Trieste, Italy}
\affiliation{$^4$ L. D. Landau Institute for Theoretical Physics,
Chernogolovka, Russia}
\affiliation{$^5$ INFN, Sezione di Trieste, Strada Costiera 11, 34151 Trieste, Italy. }
\begin{abstract}
We introduce a new measure of ergodicity, the support set $S_\varepsilon$, for random wave functions on disordered lattices. It is more sensitive than the traditional inverse participation ratios and their moments in the cases where the extended state is very sparse. We express the typical support set $S_{\varepsilon}$ in terms of the distribution function of the wave function amplitudes and illustrate the scaling of $S_{\varepsilon}\propto N^{\alpha}$ with $N$ (the lattice size) for the most general case of the multi-fractal distribution. A number of relationships between the new exponent $\alpha$ and the conventional spectrum of multi-fractal dimensions is established. These relationships are tested by numerical study of statistics of wave functions on disordered Bethe lattices. We also obtain numerically the finite-size spectrum of fractal dimensions on the Bethe lattice which shows two apparent fixed points as $N$ increases. The results allow us to conjecture that extended states on the Bethe lattice at {\it all} strengths of disorder below the localization transition are {\it non-ergodic} with a clear multifractal structure that evolves towards almost ergodic behavior in the clean limit.
\end{abstract}


\maketitle

\section{Introduction}
Anderson localization (AL) \cite{50Anderson,Anderson58}, in its broad sense, is one of the central paradigms of quantum theory. Diffusion, which is a generic asymptotic behavior of classical random walks \cite{Einstein}, is inhibited in quantum case and under certain conditions ceases to exist \cite{Anderson58}. This concerns quantum transport of non-interacting particles subject to quenched disorder as well as transport and relaxation in many-body systems. In the latter case the {\it many-body localization} (MBL) \cite{MBL} can be thought of as localization in the Fock space of Slater determinants, which play the role of lattice sites in a disordered Anderson tight-binding model. However, in contrast to a periodic $d$-dimensional lattice, the structure of Fock space is hierarchical\cite{AGKL}: a two-body interaction couples a one-particle excitation with three one-particle excitations, which in turn are coupled with five-particle excitations, etc.
This structure is reminiscent of the Bethe lattice (BL) or a regular random graph.\footnote{Although loops are present in Fock space and the connectivity might be variable from site to site, the BL is supposed to capture most of the phenomena occurring in MBL.} Interest to the problem of single particle AL on the BL \cite{Abou-Chacra, Abou-Chacra1} has recently revived \cite{Aiz-S, Aiz-War, Bir-Sem, Biroli, monthus2011anderson} largely in connection with MBL.
	It is a good approximation to consider  hierarchical lattices as trees
where any pair of sites is connected by only one path and loops are absent. Accordingly the sites being in resonance with a given site are much sparser than in ordinary $d>1$-dimensional lattices. As a result even the extended wave functions can occupy zero fraction of the BL, i.e. can be {\it non-ergodic}.
	The non-ergodic extended states on 3D lattices where loops are abundant are commonly believed \cite{Weg, AKL,  KrMut, Mir-rev} to exist but only at the critical point of the AL transition.

This paper is devoted to the analysis of the eigenstates of the Anderson model on the Bethe lattice.

 A normalized wave function $\psi(i)$  on a lattice with the total number of sites $N$ ($i=1,2,...N$) can be characterized by the moments of the inverse participation ratios $I_{q}=\sum_{i}|\psi(i)|^{2q}$~ \cite{Weg}. Suppose that $I_{q}\propto N^{-\tau(q)}$ ($I_{1}=1$) as it is in the critical point of 3D AL. Violation of the ergodicity would manifest itself by deviations of $\tau(q)$ from $(q-1)$. If at $q>1$ the ratio $D_{q}=\tau(q)/(q-1)$ is not a constant and $0<D_{q}<1$, the wave function $\psi(i)$ is called {\it multi-fractal} and is often characterized by  the {\it spectrum of fractal dimensions} $f(\alpha)$  given by the Legendre transform of $\tau_{q}$.

In this paper we introduce another measure of non-ergodicity, the {\it support set exponent}, which is more apt than  $\tau(q)$ for almost localized states. We find a general relationship of this measure with the distribution function of wave functions amplitudes $P(|\psi|^{2})$ and show how it is expressed through the spectrum of fractal dimensions $f(\alpha)$ for the generic non-ergodic multifractal states.
 Using the notion of the support set we
demonstrated that extended eigenstates of the Anderson model on the BL are
multifractal non-ergodic and extracted the spectrum of the fractal dimensions from
the numerical simulations.
\section{The support set and the amplitude distribution $P(x)$}
By the support set of a normalized wave function $\psi(i)$ with sites ordered according to $|\psi(i)|>|\psi(i+1)|\;\forall i$   we mean the set of sites $i\leq S_{\varepsilon}$ with $S_{\varepsilon}$ determined by the relation:
\be
\sum_{i=1}^{S_{\varepsilon}}|\psi_{n}(i)|^{2}\leq 1-\varepsilon<\sum_{i=1}^{S_{\varepsilon}+1}|\psi_{n}(i)|^{2}
\ee
at a given positive $\varepsilon<1$.

The normalization of $\psi(i)$ means that $S_{\varepsilon=0}=N$. However, we will be interested in the $S_{\varepsilon}(N)$-scaling in the limit when $N\rightarrow 0$ and $\varepsilon$ is arbitrary small but finite.
It is natural to call $\psi(i)$ localized if $S_{\varepsilon}$ is $N$-independent in this limit, while the scaling
$S_{\varepsilon}\propto N$ would suggest that $\psi(i)$ is both delocalized and ergodic. It is possible that $S_{\varepsilon}\rightarrow \infty$ but the ratio $S_{\varepsilon}/N\rightarrow 0$ at $N\rightarrow\infty$. In such a case we will refer to $\psi(i)$ as the {\it non-ergodic extended} wave function.

Below instead of $|\psi(i)|^{2}$ we will use variables $x(i)=N|\psi(i)|^{2}$ and define $x_{\varepsilon}=x(S_{\varepsilon})$, so that $x_{\varepsilon=0}=x(S_{\varepsilon=0})=x(N)\rightarrow 0$. Typical values of $x_{\epsilon}$ and $S_{\epsilon}$ are determined by the distribution function $P(x)$ as follows:

\bea &&\int_{x_{\varepsilon}}^{\infty}x\,P(x)\,dx=1-\varepsilon,\label{S-P-1}\\
&&\int_{x_{\varepsilon}}^{\infty}P(x)\,dx=\frac{S_{\varepsilon}}{N}\label{S-P-2}.
\eea
\section{The multifractal ansatz for $P(x)$}
Consider the amplitude distribution function $P(x)$ corresponding to the multifractal statistics (a {\it multifractal ansatz}) \cite{AKL, Mir-rev}:
\be\label{mult-anz}
P(x)=\frac{A}{N\,x}\, {\rm exp}\left[\ln N\,f(\alpha(x))\right],\;\;\;\alpha(x)=1-\frac{\ln x}{\ln N}.
\ee
with the normalization constant $A$.
\begin{figure}
\center{\includegraphics[width=0.7\linewidth]{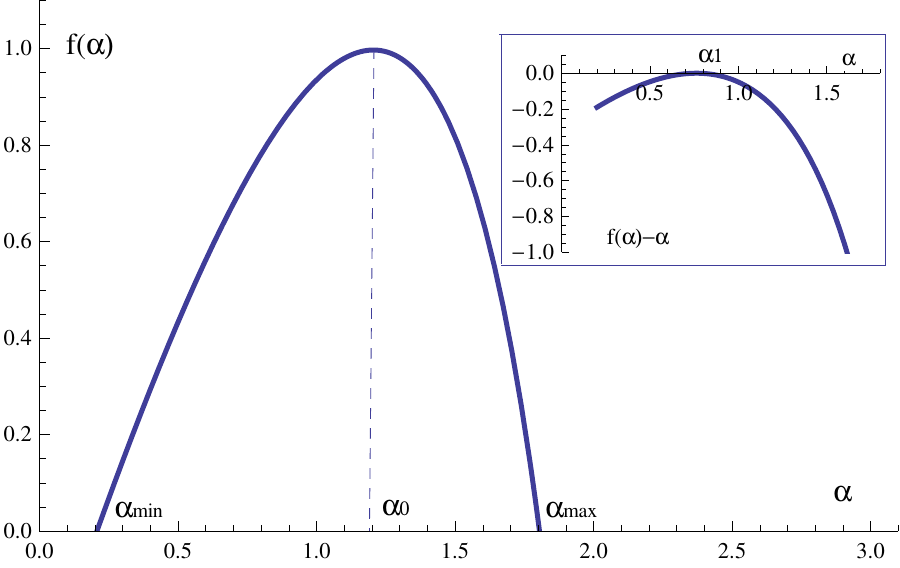}}
\caption{(Color online)
 Non-singular function $f(\alpha)$ obeying the symmetry Eq.(\ref{sym-f-alpha}).  In the inset: a generic function $f(\alpha)-\alpha$. } \label{Fig:f-alpha-gen}
\end{figure}
 The spectrum of fractal dimensions $f(\alpha)$ defined for $\alpha>0$  (i)  is a convex function, and (ii) its maximal value equals to 1: $f_{max}=f(\alpha_{0})=1$. Besides that, for extended states  a number constraints for $f(\alpha)$ follow from the symmetry of the distribution  $F(\rho)$ of the local density of states (LDoS) $\rho(\epsilon, i)=\sum_{n}|\psi_{n}(i)|^{2}\,\frac{\Gamma_{n}}{\pi\,[(\epsilon-\epsilon_{n})^{2}+\Gamma_{n}^{2}]}$:
\be\label{MF-sym}
F(\rho)=\rho^{-3}\,F(\rho^{-1}).
\ee
This symmetry was first established for the localized states in strictly one-dimensional system with random potential \cite{AlPrig}, and later  derived from the nonlinear supersymmetric sigma-model \cite{Mir-Fyod-sym, Mir-Fyod-sym1}.
The level widths $\Gamma_{n}$  do not fluctuate much if the states are extended. Accordingly fluctuations of $\rho$  are mostly due to the fluctuations of the wave-function amplitude $|\psi_{n}(i)|^{2}$, i.e. of x.
Therefore P(x) should obey the symmetry relation Eq.(\ref{MF-sym}) as well as $F(\rho)$. When applied to the distribution Eq.(\ref{mult-anz}) the symmetry Eq.(\ref{MF-sym}) imposes a functional constraint:
\be\label{sym-f-alpha}
f(1+\alpha)=f(1-\alpha)+\alpha.
\ee
Equation Eq.(\ref{sym-f-alpha}) is valid only for extended states. Indeed, the fluctuations of LDoS $\rho$ of the localized states are dominated by the fluctuations of $\Gamma_{n}$ and there is no reason for Eq.(\ref{MF-sym}) to hold for $P(x)$.

It follows from Eq.(\ref{sym-f-alpha})  that $f'(1)=\frac{1}{2}$, i.e. the only power-law distribution $P(x)$ (linear function $f(\alpha)$ ) compatible with Eq.(\ref{sym-f-alpha}) is:
\be\label{power-law}
P(x)= A x^{-\frac{3}{2}}.
\ee
As we will see, $P(x)$ acquires the form Eq.(\ref{power-law}) at the point of the Anderson transition on the BL.

The function $f(\alpha)$ reaches its maximum $f(\alpha_{0})=1$ at $\alpha=\alpha_{0}$, i.e. $f'(\alpha_{0})=0$.
Substituting $\alpha=\alpha_{0}-1$ into Eq. (\ref{sym-f-alpha}) and its derivative over $\alpha$  we find that for
$\alpha_{1}=2-\alpha_{0}$ holds:

\be\label{alpha-1}
f(\alpha_{1})=\alpha_{1},\;\;\;{\rm and}\;\;\;f'(\alpha_{1})=1.
\ee
 The straight line $f(\alpha)=\alpha$  is thus a tangential to $f(\alpha)$ at the point $\alpha=\alpha_{1}=2-\alpha_{0}$. Therefore convexity of $f(\alpha)$  implies that $\alpha_{1}<1$  and $\alpha_{0}>1$,  since $0<f'(1)=1/2<1$.
  In a similar way one may prove that the solutions $\alpha_{q}$ to the equations $f'(\alpha_{q})=q$
are ordered ($\alpha_{q}>\alpha_{q+1}$) and:
\be\label{alpha-q}
0<q\alpha_{q}-f(\alpha_{q})<\alpha_{q-1}.
\ee
\subsection{The support set and the function $f(\alpha)$}
For $P(x)$ given by Eq.(\ref{mult-anz})  one obtains from  Eqs.(\ref{S-P-1}),(\ref{S-P-2}):
\be\label{set-f}
S_{\varepsilon}=\varepsilon\;\frac{\int_{0}^{\alpha_{\varepsilon}}d\alpha\;e^{\ln N\,f(\alpha)}}{\int_{\alpha_{\varepsilon}}^{\infty}d\alpha\,e^{\ln N\,[f(\alpha)-\alpha]}},
\ee
where $\alpha_{\varepsilon}$ is a solution to the equation:
\be\label{alpha-eps}
\frac{\int_{\alpha_{\varepsilon}}^{\infty}d\alpha\;e^{\ln N\,[f(\alpha)-\alpha]}}{\int_{0}^{\infty}d\alpha\,e^{\ln N\,[f(\alpha)-\alpha]}}=\varepsilon.
\ee
At large $\ln N$ all the integrals in Eqs.(\ref{set-f}),(\ref{alpha-eps}) are dominated by $\alpha$ in the vicinity of maxima $\alpha_{0}$ and $\alpha_{1}$ of the functions $f(\alpha)$ and $f(\alpha)-\alpha$, respectively, or the end points of the integration domain and can be done by the saddle-point or the end-point approximations.

\subsection{Support set for a non-singular function $f(\alpha)$}
Consider now a generic {\it non-singular} function $f(\alpha)$ obeying the functional constraint Eq.(\ref{sym-f-alpha}). In this case
the maximal values 1 and 0 of $f(\alpha)$ and $f(\alpha)-\alpha$  are reached at $\alpha=\alpha_{0}$ and $\alpha=\alpha_{1}$ such that $\alpha_{min}<\alpha_{1}<1<\alpha_{0}<\alpha_{max}$ (see Fig.~\ref{Fig:f-alpha-gen}).

It follows from these observations that
\be
\frac{\ln(C/\varepsilon)}{\ln N}= \alpha_{\varepsilon}-f(\alpha_{\varepsilon}),\;\;\;(\alpha_{\varepsilon}>\alpha_{1}),
\ee
%
%
where $C=\sqrt{|f''(\alpha_{1})|/(2\pi\ln N)}(1-f'(\alpha_{\varepsilon}))^{-1}$ is weakly $N$-dependent.
The expression for the number of sites in the support set $S_{\varepsilon}$ depends on whether $\alpha_{\varepsilon}$ is smaller or greater than $\alpha_{0}$ corresponding to the maximum of $f(\alpha)$:
\bea
S_{\varepsilon}&=& C_{1}\,\varepsilon\;N^{\alpha_{\varepsilon}},\;\;\;  \ln N > \frac{\ln C/\varepsilon}{\alpha_{0}-1},\;\;\;(\alpha_{\varepsilon}<\alpha_{0}),\\
S_{\varepsilon}&=& C_{2}\,N,\;\;\;\; \ln N < \frac{\ln C/\varepsilon}{\alpha_{0}-1},\;\;\;(\alpha_{\varepsilon}>\alpha_{0}),
\eea
where $C_{1}=(1-f'(\alpha_{\varepsilon}))\,|f'(\alpha_{\varepsilon})|^{-1}$, $C_{2}=\sqrt{f''(\alpha_{1})/f''(\alpha_{0})}$.

Thus an $\varepsilon$-dependent correlation length
\be
\zeta_{\varepsilon}=\ln(C/\varepsilon)/(\alpha_{0}-1)
\ee
develops such that for $\ln N<\zeta_{\varepsilon}$ the support set $S_{\varepsilon}\propto N$ is like in the ergodic metal. This correlation length diverges in the limit of weak multifractality when $\alpha_{0}\rightarrow 1$. As $\ln N$ increases above $\zeta_{\varepsilon}$, the non-ergodic nature of the support set exhibits itself, and in the limit $\ln N\rightarrow\infty$ one has:
\be\label{sup-set-non-erg}
S_{\varepsilon}=\varepsilon\, N^{\alpha_{1}}\,e^{\sqrt{\frac{2\ln(1/4\pi\varepsilon)}{|f''(\alpha_{1})|}\ln N}}\,\sqrt{\frac{2\ln(1/4\pi\varepsilon)\,|f''(\alpha_{1})|}{\ln N}},
\ee
where $\alpha_{1}<1$  satisfies the equation $f(\alpha_{1})=\alpha_{1}$ (Fig.~\ref{Fig:f-alpha-gen}).

\subsection{The support set and the participation ratio}
The moments of inverse participation ratio $I_{q}$ are expressed in terms of $f(\alpha)$ as follows:
\be\label{mom-IPR}
I_{q}=\frac{\int_{0}^{\infty}d\alpha\;e^{\ln N\,[f(\alpha)-q\alpha]}}{\int_{0}^{\infty}d\alpha\,e^{\ln N\,[f(\alpha)-\alpha]}}
\ee
Note that the scaling with $N$ of the support set in Eq.(\ref{sup-set-non-erg}) is different from that of the inverse moment $I_{2}$. Indeed, Eq.(\ref{mom-IPR}) results in:
\be\label{inv-I-2}
I_{q}^{-1}\propto N^{D_{q}},\;\;\;\;
D_{q}\equiv \min_\alpha\{ q\alpha-f(\alpha)\}
\ee
From Eq.(\ref{alpha-q}) at $q=2$ it follows that $D_{2}< \alpha_{1}$, which means that the participation ratio $I_{2}^{-1}$ is sparser and stronger fluctuating than the support set.
\subsection{Support set in the "frozen" phase.}
\begin{figure}
\center{\includegraphics[width=0.6\linewidth]{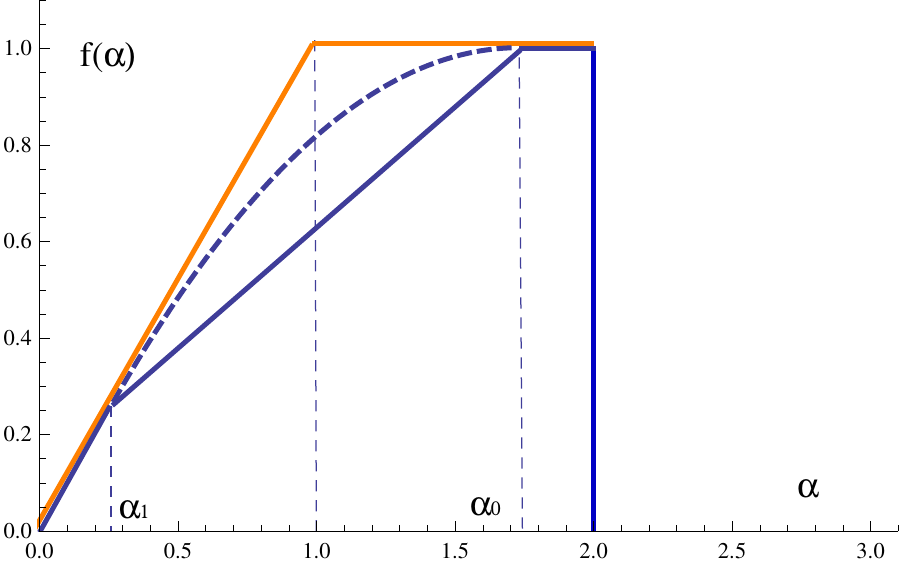}}
\caption{(Color online)
The sketch of the function $f(\alpha)$ in the frozen extended phase (blue) and at the freezing transition (orange). For such $f(\alpha)$ all the fractal dimensions $D_{q}=0$ for $q>1$ while the support set $S_{\varepsilon}\propto N^{\alpha_{1}}$ has a scaling exponent $\alpha_{1}>0$.} \label{Fig:plot-f-freez}
\end{figure}
 An interesting situation appears when the minimal value of $2\alpha-f(\alpha)$ is achieved at $\alpha=0$ and it is zero. This corresponds to a special phase which arises in some systems below the so called {\it freezing transition} \cite{deChamon, CLeDous} where all $D_{q}$ with $q>1$ in Eq.(\ref{inv-I-2}) are zero, while $\alpha_{1}$ in Eq.(\ref{sup-set-non-erg}) may be non-zero.
This is only possible if $f(\alpha)\equiv \alpha$ for $\alpha<\alpha_{1}$. Then the  symmetry Eq.(\ref{sym-f-alpha}) requires that $f(\alpha)\equiv 1$ for $\alpha_{0}=2-\alpha_{1}<\alpha<2$ followed by an abrupt termination at $\alpha=2$. For $\alpha_{1}<\alpha<\alpha_{0}$ the $f(\alpha)$ may be a smooth function with the derivative decreasing  from 1 to 0 but the symmetry Eq.(\ref{sym-f-alpha}) allows it also to be a linear function with a fixed slope $1/2$.  For such  $f(\alpha)$ {\it all} the moments with $q>1$ depend slowly on $\ln N$:  $I_{q}\propto 1/(1+\frac{\alpha_{1}}{2}\,\ln N)$ and are reminiscent of those in the insulating phase $I^{\rm ins}_{q}={\rm const}$. In contrast to the moments, the support set  $S_{\epsilon}\propto N^{\alpha_{1}}/[\epsilon\,(1+\frac{\alpha_{1}}{2}\,\ln N)^{2}]$ is almost a power law with the exponent $\alpha_{1}$, and thus it is very different from that in the insulator $S^{\rm ins}_{\epsilon}\propto {\rm const}$. This is the best example illustrating the sensitivity of the support set as a measure of ergodicity of the wave function.
\section{Function $f(\alpha)$ in the insulator on the Bethe lattice.}
Consider the Anderson Hamiltonian on the Bethe lattice
\begin{equation}
H=-t\sum_{<ij>}(c^\dag_i c_j+\mbox{h.c.})+\sum_i \varepsilon_i c^\dag_i c_i.
\end{equation}
We will now show that the locator expansion in the forward scattering, or  directed polymer, \cite{MedinaKardar,Derrida}, approximation leads to the following expression for the distribution function $P(x)$ in the insulator on the Bethe lattice with the large branching number $K\gg 1$ and the disorder parameter $W$ (for the box distribution $\varepsilon_{i}\in[-W/2,W/2]$ and the hopping integral $t=1$):

The application of the forward scattering approximation \cite{MedinaKardar} to the Schroedinger equation on the BL \cite{AGKL} is particularly easy since there is only one shortest path between any two given points (as a counterexample on $d$-dimensional hypercubes the reader might think of the case of two points opposite on the diagonal). Starting in the deep localized region, we consider a wave function $\psi$ centered at the site $0$ with energy $\varepsilon_0$. We have that, the amplitude at a site $i$, in the forward approximation
\begin{equation}
\psi(i)=\prod_{j\in p}\frac{t}{\varepsilon_0-\varepsilon_j}
\end{equation}
where $p$ is the shortest path from $0$ to $i$, which length is $n$ (from now on we will set $t=1$, the scale of energies). It is convenient to pass to $x_n=N\psi(i)^2$ and study the distribution of $\ln x_n$, as this is a sum of i.i.d.\ random variables distributed with one-parameter probability density $\tilde\rho(\varepsilon_i)=\frac{1}{W}\rho(\varepsilon_i/W)$. For simplicity we assume $\varepsilon_0=0$ and defining dimensionless $Z=(2/W)^2$ and $y_j=((W/2)/\varepsilon_j)^2$:
\begin{equation}
\ln x_n=\ln N+n \ln ((2/W)^2)+\sum_{j=1}^n\ln y_j.
\end{equation}
For the case of the box distribution $\rho(e)=2^{-1}\Theta_{[-1,1]}(e)$ and therefore $y_j>1$ and $\xi_n=\ln x_n-n\ln (N^{1/n}(2/W)^2)=\sum_{j}\ln y_j>0$ (in the case of more general $\rho(e)$ one should resort to Fourier transform but this technicality does not change the calculations substantially). We find
\begin{equation}
\label{eq:supmatpy}
p(y)=\frac{1}{2y^{3/2}}\theta(y-1).
\end{equation}
The power law tail at large $y$ is a common feature of any distribution and arises from the divergence of the denominators, which inhibits the existence of the average of $y$.

As usual, the Laplace transform of the sum of i.i.d.\ variables is the $n$-th power of the Laplace transform of that of a single variable which in this case is:
\begin{eqnarray}
R(s)&=&\int_0^\infty d\ln y\ e^{-s\ln y} p(\ln y)=\nonumber\\
&=&\int_1^\infty dy\ y^{-s}\frac{1}{2y^{3/2}}=\frac{1}{1+2s}.
\end{eqnarray}
So by taking the $n$-th power and inverting the Laplace transform we have formally:
\begin{equation}
P_n(\xi_n)=\int_B\frac{ds}{2\pi i} e^{s\xi_n}R(s)^n,
\end{equation}
where the Bromwich path $B$ passes to the right of the only singularity of the integrand, ($s=-1/2$ in the case of the box distribution).

Therefore the distribution of the $x_n=N(2/W)^{2n}e^{\xi_n}$ is
\begin{equation}
P_n(x_n)=\frac{1}{x_n}\left. P_n(\xi_n)\right|_{\xi_n=\ln x_n-\ln N-2n\ln (W/2)}
\end{equation}
so
\begin{equation}
\label{eq:supmatPn}
P_n(x_n)=\frac{1}{x_n}\int_B\frac{ds}{2\pi i}\left(\frac{x_n}{N}\right)^s(W/2)^{-2ns}R(s)^n.
\end{equation}
We find now the probability distribution $P(x)$ by summing over the events that the given observation site $i$ belongs to the $n$-th generation:
\begin{equation}
\label{eq:supmatP}
P(x)=\sum_{n=1}^{\ln N/\ln K}\frac{K^{n-1}(K+1)}{N}P_n(x).
\end{equation}
The sum over $n$, if convergent, can be extended to $n=\infty$ and exchanging the integral and sum we get a geometric series. In order to ensure convergence the Bromwich path has to be shifted in between $s_-$ and $s_+$ as discussed in the main text. The result of the geometric series is (with a further redefinition $2s+1\to s$):
\be \label{P-k-forward}
P(x)=\frac{1}{2N^{\frac{1}{2}}\,x^{\frac{3}{2}}}\int_{B}\frac{d s}{2\pi i}\,\frac{s\,(x/N)^{\frac{s}{2}}}{s-K\,(W/2)^{(s-1)}}.
\ee
The contour $B\in (r-i\infty, r+i\infty)$ is parallel to the imaginary axis and crosses the real axis at $s_{-}<r < s_{+}$, where $s_{\pm}$ are the larger and the smaller real root of the equation
\be\label{pole-eq}
s=K\,(W/2)^{(s-1)}.
\ee
One can check that this equation has real roots if and only if $W\geq W_{c}$, where:
\be\label{AT-point}
W_{c}/2=e\,K\,\ln(W_{c}/2)\approx e\,K\,\ln(e\,K),
\ee
which  coincides exactly with the critical disorder given by Eq.(84) of the seminal work \cite{Anderson58} of Anderson (see also the ``upper limit critical condition'' of Ref.\cite{Abou-Chacra, Abou-Chacra1}). Note that for a normalized wave functions on a lattice holds $x=N|\psi_{n}|^{2}<N$ and thus $x<(W_{c}/2)^{2}\,N < (W/2)^{2}\,N$. The last condition allows to bend the contour $B$ to enclose the part of real axis $\Re s>r$ which includes the pole $s=s_{+}$ and the other ones in complex conjugate pairs at $\Re{s}\gg \Re{s}_{+}$.

Then Eq.(\ref{P-k-forward}) gives immediately the leading order as a power-law
\be\label{power-law-s-plus}
P(x)= N^{-(1-\beta)}\,x^{-(1+\beta)}+o(N^{-(1-\beta)}),\;\;\;\;\beta=\frac{1}{2}-\frac{s^{+}}{2}.
\ee

It is now tempting to pause for a moment and to draw a parallel between the form of Eq.s\ (\ref{eq:supmatPn}) and (\ref{eq:supmatP}) and the resummation of leading logarithms in renormalization group. In fact, each of the terms in Eq.\ (\ref{eq:supmatPn}) has a simple pole at $s=-1/2$ which gives the ``bare" power law $x^{-3/2}$ which comes from the divergence of the denominators in the very same way as that in Eq.\ (\ref{eq:supmatpy}). However by summing over every generation $n$ we obtain an effective power law with a different, $W$-dependent power $\beta$. The Laplace-Mellin transform is again a powerful tool to extract such power laws.

This procedure fails when $W$ becomes small enough such that the geometric series does not converge for any value of $s$ anymore, as testified by the poles $s_{\pm}$ moving out of the real axis. However, the failure of this approximation occurs most likely before (although it is exact in the large $K$ limit) due to the failure of the forward approximation for the amplitudes. Physically this is equivalent to neglecting the corrections to the real part of the self-energy of the sites $j$ on the path from 0 to $i$ and once these are taken into account we have to consider the possibility that a resonance occurs with probability higher than the ``bare" $p(y)$ predicts. One can correct this in perturbation theory in $1/\ln K$ and show that the power-law form is preserved. We leave this for a future publication as a new set of tools is needed to discuss the perturbation theory in $1/\ln K$.

To connect with the previous discussion notice that (\ref{power-law-s-plus}) is described by Eq.(\ref{mult-anz}) with
\be\label{f-ins}
f(\alpha)=\beta\,\alpha,\;\;\;\alpha<1/\beta.
\ee
which should be terminated at $\alpha_{0}=\beta^{-1}$.\footnote{Power law distribution of the wave-function coefficients in the many-body localized region of a Heisenberg chain with random fields was already observed in \cite{DeLuca}} This corresponds to the truncation of the distribution function $P(x)$ in Eq.(\ref{power-law-s-plus}) at $x_{{\rm min}}=N^{1-\frac{1}{\beta}}$.

At a very large disorder $W\rightarrow\infty$ the larger solution to Eq.(\ref{pole-eq}) $s^{+}\rightarrow 1$. Correspondingly $\beta\rightarrow 0$ at $W\rightarrow\infty$:
\be\label{beta-large-W}
\beta\approx \frac{\ln K}{2\ln(W/2)},\;\;\mbox{for}\;\;\ln(W/2)\gg \ln K.
\ee
At $W\rightarrow W_{c}$ one finds $s_{+}=s_{-}=\frac{1}{\ln(W_{c}/2)}\sim \frac{1}{\ln K}$. Thus in the $K\rightarrow\infty$ limit, in which our approach  is valid,
\be\label{beta-cr}
\beta_{c}=\frac{1}{2},
\ee
in agreement with \cite{Abou-Chacra}.
Eq.(\ref{beta-cr}) follows from the symmetry of $f(\alpha)$.
Indeed, Eq.(\ref{power-law-s-plus}) should be valid in insulator down to the critical point. On the other hand, the critical distribution function should be a limiting case of $P(x)$ for extended states and thus it should obey the symmetry relation Eq.(\ref{MF-sym}). The only power-law distribution that obeys Eq.(\ref{MF-sym}) is Eq.(\ref{power-law}) which corresponds to $\beta=\frac{1}{2}$. A similar statement has been done in Ref.\cite{Abou-Chacra} for the power-law distribution of the imaginary part of the Green's function based on a symmetry of the equations (Eq.(6.8) in Ref.\cite{Abou-Chacra}) which is similar to Eq.(\ref{MF-sym}).

Now plugging Eq.(\ref{f-ins}) into Eqs.(\ref{set-f}),(\ref{alpha-eps}) one obtains:
\bea
S_{\varepsilon}= (\beta^{-1}-1)\,
\varepsilon^{-\frac{\beta}{1-\beta}},\;\;\;
\varepsilon\gg N^{-(\beta^{-1}-1)}.
\eea
One can see that the number of sites in the support set is independent of $N$ provided that $N\gg \varepsilon^{-\frac{1}{\beta^{-1}-1}}$. This is the behavior typical of insulator. The peculiarity of the Bethe lattice is that the same behavior holds also at the Anderson transition point.

\begin{figure}
\center{\includegraphics[width=0.6\linewidth]{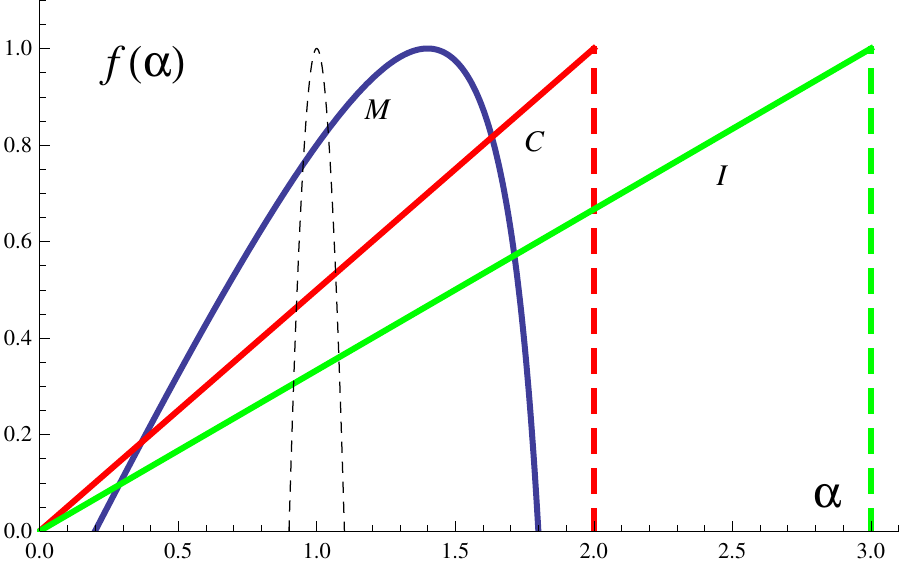}}
\caption{(Color online)
The sketch of the function $f(\alpha)$ on the Bethe lattice with $K\gg 1$: (I) for the localized phase (green);(C) for the Anderson transition point (red); (M) for the non-ergodic delocalized phase (blue). The black dashed curve corresponds to the almost ergodic delocalized phase. For the localized and the critical state $f(0)=0$ ($\alpha_{min}=0$). For extended, non-ergodic states $\alpha_{min}>0$.} \label{Fig:plot-f-I-C-M}
\end{figure}
\section{Scenario for a non-ergodic metal phase on the Bethe lattice.} First of all let us note that the description in terms of the multifractal ansatz Eq.(\ref{mult-anz}) is only valid when the wave functions are non-ergodic. Since it is essentially proven that only two types of spectra exist on the Bethe lattice \cite{Aiz-War}, it is most likely that there is only one transition from the localized to the delocalized phase and that this corresponds to the transition predicted by the locator expansion of Ref.\cite{Abou-Chacra}. However, it is not excluded that the delocalized phase is {\it non-ergodic} at any strength of disorder with the corresponding function $f(\alpha)$ that experiences a crossover from the critical triangle form at $W=W_{c}$ to the almost ergodic parabolic form at weak disorder.
The corresponding sketch for $f(\alpha)$ is shown in Fig.~\ref{Fig:plot-f-I-C-M}. We stress that it has been rigorously proven\cite{aizenman2012absolutely} that the imaginary part of the self-energy
can not have long tails throughout the delocalized region.\footnote{We thank V. Bapst for putting this fact under our attention.} This quantity is definitely related to the wave-function amplitude distribution, however, an averaging procedure
on a small but fixed energy window is needed, which can actually eliminate the long tail of $P(x)$. In the next section we show how this happens at small $x$. At large $x$ it is fundamental that within the multi-fractal ansatz the effective cut off is provided by $\alpha_{min}>0$ (see Figs.\ref{Fig:f-alpha-gen},\ref{Fig:plot-f-I-C-M}) which is a hallmark of an extended state.

It is also possible that the statistics of the energy levels in the delocalized region is neither Poisson nor Wigner-Dyson,
as a result of strong correlations between neighboring levels which exists despite them being fractals. This correlation is also fundamental
to produce a non-trivial self-energy distribution and is captured by Chalker's scaling exponent $\mu$, \cite{Chalker}. We plan to analyze it in details in a forthcoming publication.

\section{The amplitude distribution function $P(x)$ at small $x$.}

The distributions $P(x)$ we have discussed up to now, in the general form of Eq.(\ref{mult-anz}), describe the smooth envelope $\psi_{en}$ of the fast oscillating wave function $\psi$. This is, for some values of $x$, very much different from the numerically obtained distribution function ${\cal P}(N|\psi^{2}|)$ of the values of the wave function. For example, according to the multifractal ansatz, there is always a minimal statistically relevant $|\psi_{en}|^{2}=N^{-\alpha_{{\rm max}}}$ while ${\cal P}(N|\psi^{2}|)$ does not have this feature as $\psi$ can be arbitrarily close to 0 for finite $N$, due to interference effects.

We shall now devise a method, alternative to the existing ones and better suited for the Bethe lattice, to recover $P_{en}(x_{en})$ and from this, $f(\alpha)$. The numerical estimation of the fractal spectrum encoded in the function $f(\alpha)$ is usually a complicated task due to the fluctuation of the eigenstates on the scale of the lattice length. This fact can be seen from the function $P(x)$ that always presents a square-root behavior $x^{-1/2}$ at small $x$. Other approaches are known to overcome this difficulty usually based on a real-space renormalization procedure at large wavelengths, usually called \textit{box counting} \cite{rodriguez2009optimisation}.
In our case, this procedure clashes with the exponential growth of the BL so that even for the largest sizes we can numerically achieve the spatial extension of the system remains rather small (the diameter of our largest system counts about 16 nodes).
For this reason we follow a different method. It is based on the assumption that the variable $x_{ED} = N|\psi|^2$, coming from exact diagonalization, can be written in the form
\begin{equation}
 \label{xdec}
 x_{ED} = x_{en} x_{GOE}
\end{equation}
where $x_{GOE}$ corresponds to fast oscillations in the Gaussian Orthogonal Ensemble with the distribution function $P_{GOE}(x_{GOE})=e^{-x_{GOE}/2}/\sqrt{2\pi x_{GOE}}$. Switching to logarithmic variables $\ln x=\ln x_{en}+\ln x_{GOE}$ one can obtain $P_{en}(x_{en})$ from the convolution of distribution functions ${\cal P}(\ln x)=P_{en}(\ln x_{en})*P_{GOE}(\ln x_{GOE})$ obtained numerically. At small $x$ where the distribution of envelope  $P_{en}(x)$ rapidly vanished, the distribution of the total wave function amplitude ${\cal P}(x)\propto x^{-1/2}$ is a power-law.
\begin{figure}
\center{\includegraphics[width=0.8\linewidth]{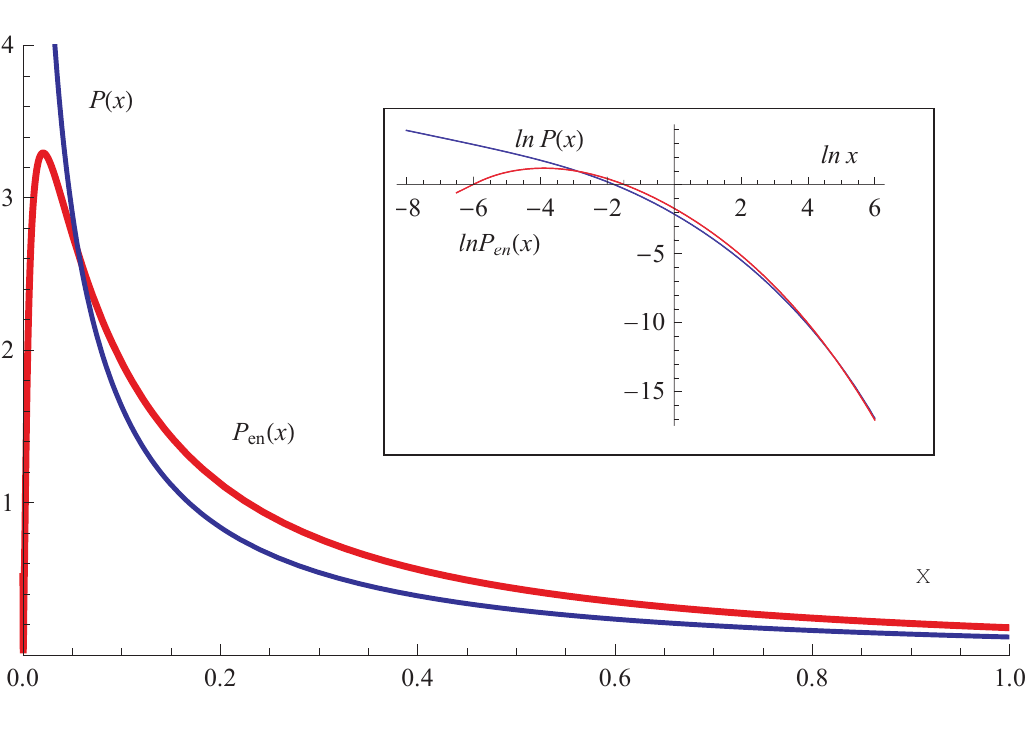}}
\caption{(Color online)
The PDF of the wave function amplitude ${\cal P}(x)$ (blue) and the PDF of the corresponding smooth envelope $P_{en}(x)$ (red) on the Bethe lattice with $K=2$ and N=16000 at disorder strength $W=7.5$.} \label{Penv}
\end{figure}

The distribution of $\ln x$ can be obtained numerically by binning the eigenvectors, while the distribution of $\ln x_{GOE}$ is explicitly known. In this way, the distribution of $\ln x_{en}$ can be obtained efficiently inverting the convolutions with the help of fast-Fourier transform
\begin{equation}
 \label{fft}
 \mathcal{Q}_{en}(k) = \frac{\mathcal{Q}_{ED}(k)}{\mathcal{Q}_{GOE} (k)}=\frac{2^{-i k-a^{2}k^{2}}\Gamma\left(\frac{1}{2}\right)}{\Gamma\left(\frac{1}{2}+ i k\right)}\,\mathcal{Q}_{ED}(k)
\end{equation}
where $\mathcal{Q}(k)$ generically indicates the Fourier transforms of the distributions $\mathcal{P}(\ln x)$.
The only difficulty comes from the fact that the data of the distribution of $\ln x$ are affected by errors which spoil the behavior of the Fourier transform at large $k$. The result is that the right-hand side of \eqref{fft} explodes at large $k$, making the inversion rather unstable.
To avoid this problem, we smoothed out the data of $\mathcal{P}_{ED}(\ln x)$ with a Gaussian kernel with a characteristic width $a$. This adds an additional Gaussian factor $e^{- a^2 k^2}$ in the right-hand side of  \eqref{fft} which ensures convergence. Ideally, the original equation is recovered only when the width of the Gaussian kernel $a$ is sent to zero. However, we checked that
the results are sufficiently robust when the width is decreased until the numerical errors become too relevant ($a^2 \gtrsim 0.1$).

\section{Numerics on the Bethe lattice.}  By exact numerical diagonalization of the Anderson Hamiltonian on the Bethe lattice we study as a function of the strength of disorder W the following statistics: $({\bf i})$ the exponent $\alpha$ of the typical support set $\alpha=\ln S_{\varepsilon,{\rm typ}}/\ln N=\langle \ln S_{\varepsilon} \rangle/\ln N$; $({\bf ii})$  the fractal dimension $D_{2}$;
  $({\bf iii})$ the exponent $\gamma(W)$ in $y=N^{\gamma}\,|\psi|^{2}$ that enables the best collapse on the same universal function of the bulk of the  distribution function ${\cal P}(y)$ for different $N$.
  The diagonalization procedure is performed by the Lanczos technique which we used to compute exactly a fixed number of eigenfunctions ($\simeq 100$) around the middle of the band.

\begin{figure}
\center{\includegraphics[width=0.7\linewidth]{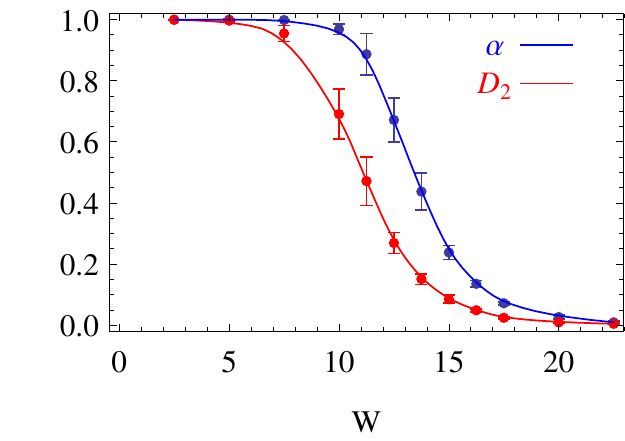}}
\caption{(Color online)
 Exponents $D_{2}(W)$ and  $\alpha(W)$ for the Bethe lattice with $K=2$ and $N=16000$. The strong inequality $\alpha>D_{2}$ that follows from the multifractal picture is well fulfilled. At the same time, $\alpha=D_{2}$ in the insulator, critical, and extended ergodic phase in the limit $\ln N\rightarrow \infty$.} \label{Fig:D-2-alpha}
\end{figure}

In Fig.~\ref{Fig:D-2-alpha} we present as a function of disorder strength both the exponent $\alpha$ of the typical support set and the exponent $D_{2}$ which characterizes the moment $I_{2}$.
Both of them show a crossover from localization $\alpha=D_{2}=0$ to the extended ergodic behavior $\alpha=D_{2}=1$. However, the support set exponent $\alpha$ is distinctly larger than $D_{2}$ in full accordance with the multifractal picture. At the same time, $\alpha=D_{2}$ in the extended ergodic, critical and the localized phases in the limit $\ln N\rightarrow\infty$. This suggests that the same equality is likely to hold also for a finite  $\ln N$, were the extended non-ergodic phase absent.

According to Eq.(\ref{sup-set-non-erg}), $\alpha(W)=\alpha_{1}$ (see Fig.~\ref{Fig:f-alpha-gen}) in the large $\ln N$ limit. The exponent $\gamma$ represents the typical scaling of the most abundant values of $|\psi|^{2}\propto N^{-\gamma}$ close to the maximum of $f(\alpha)$.
We conclude therefore that $\gamma(W)=\alpha_{0}$ in the limit of large $\ln N$. Thus from the numerics for $\alpha(W)$ and $\gamma(W)$ we can find the two principle parameters $\alpha_{1}$  and $\alpha_{0}$ of $f(\alpha)$. In the region of extended states they should be connected by a simple relation $\alpha_{0}=2-\alpha_{1}$ that follows from the symmetry Eq.(\ref{sym-f-alpha}). Thus if the description in terms of the multifractal ansatz Eq.(\ref{mult-anz}) is valid we expect:
\be\label{gamma-alpha}
\gamma(W)=2-\alpha(W), \;\;\;\;W<W_{c}.
\ee
Remarkably, this relation is fulfilled within the error bars (see Fig.~\ref{Fig:abg}):
\begin{figure}
\center{\includegraphics[width=0.7\linewidth]{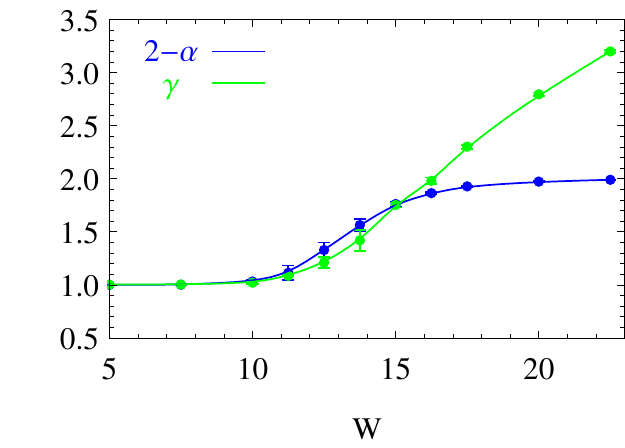}}
\caption{(Color online)
 Numerical results for $2-\alpha(W)$ and $\gamma(W)$   on the Bethe lattice with $K=2$ for $N=2,4,8,16,32\times 10^3$. In the region of extended states the relation $\gamma=2-\alpha$ is fairly well fulfilled with the smooth crossover from $\gamma\approx2$ at the localization transition to $\gamma\rightarrow 1$ at weak disorder.} \label{Fig:abg}
\end{figure}
\begin{figure}
\center{\includegraphics[width=0.7\linewidth]{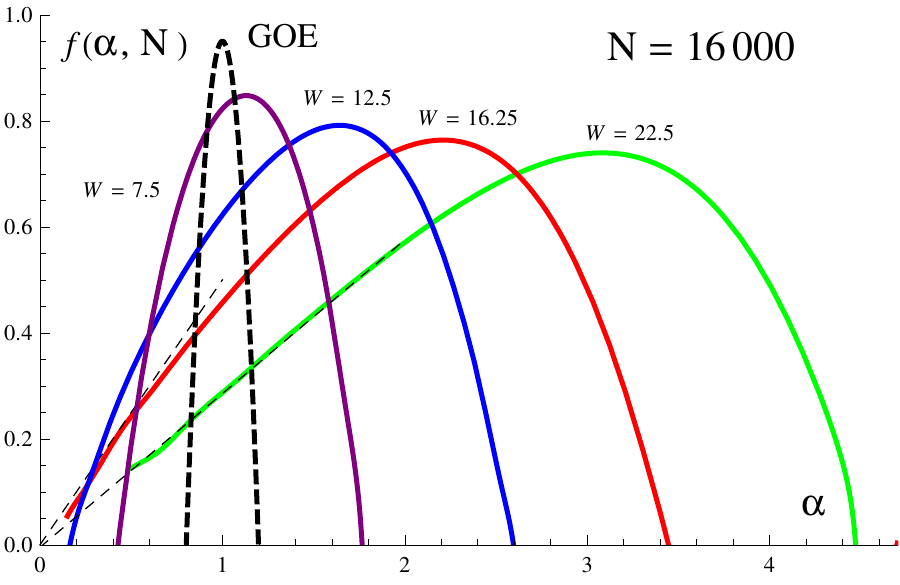}}
\caption{(Color online)
 Numerical results for $f(\alpha,N)$ on the Bethe lattice with $K=2$ at $N=16000$ for different values of disorder $W$. The dashed straight lines show the initial slope $\beta$ for the localized and close-to-critical states which is smaller than (for localized states with $\alpha_{min}=0$) or equal to (for near critical states with the smallest non-zero $\alpha_{\min}$) 1/2.} \label{Fig:f-num}
\end{figure}

\begin{figure}
\center{\includegraphics[width=0.7\linewidth]{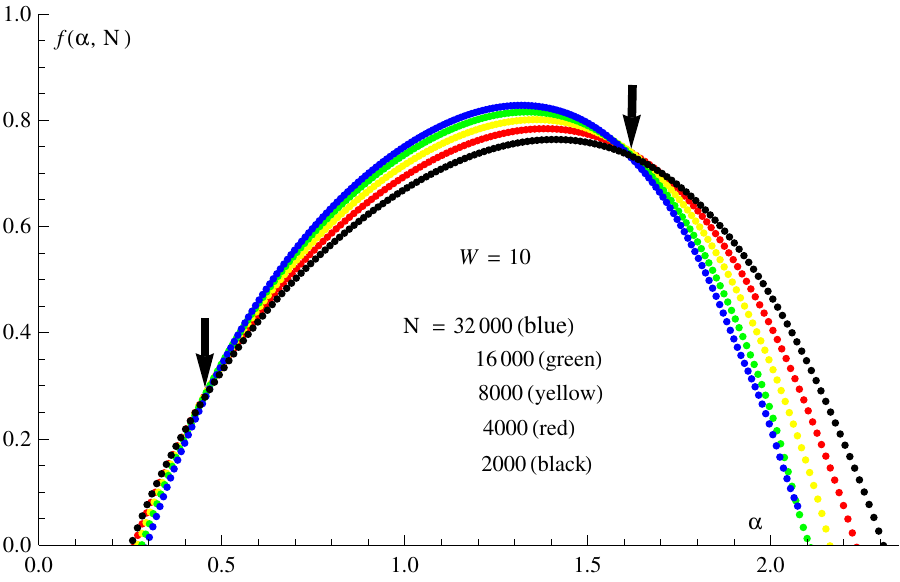}}
\caption{(Color online)
$N$-dependence of $f(\alpha,N)$ on the Bethe lattice with $K=2$ for $N=2,4,8,16,32\times 10^3$ (from black to blue in ascending order) in the extended phase $W=10$. The plots for different $N$ show  apparent fixed points at $\alpha\approx 0.5$ and $\alpha\approx 1.6$ indicated by arrows. Similar fixed points with $W$-dependent positions are seen at any strength of disorder studied.  This makes it unlikely that $f(\alpha,N)$ approaches the GOE limit $\alpha_{min}\approx\alpha_{max}\approx 1$ at $\ln N\rightarrow\infty$.} \label{Fig:f-N-dep}
\end{figure}

Finally, in Fig.~\ref{Fig:f-num} and Fig.~\ref{Fig:f-N-dep} we present
\be\label{f-alpha-N}
f(\alpha,N)=\ln [N x P_{en}(x)]/\ln N
\ee
as a function of $\alpha=1-\ln x/\ln N$. If the multifractal ansatz Eq.(\ref{mult-anz}) applies, $f(\alpha,N)\rightarrow f(\alpha)$ in the limit $\ln N\rightarrow\infty$.
Indeed, Fig.~\ref{Fig:f-num} obtained on the Bethe lattice with $K=2$ is visibly similar to the sketch Fig.~\ref{Fig:plot-f-I-C-M} drawn using the arguments valid at $K\gg 1$.

\begin{figure}
\center{\includegraphics[width=0.7\linewidth]{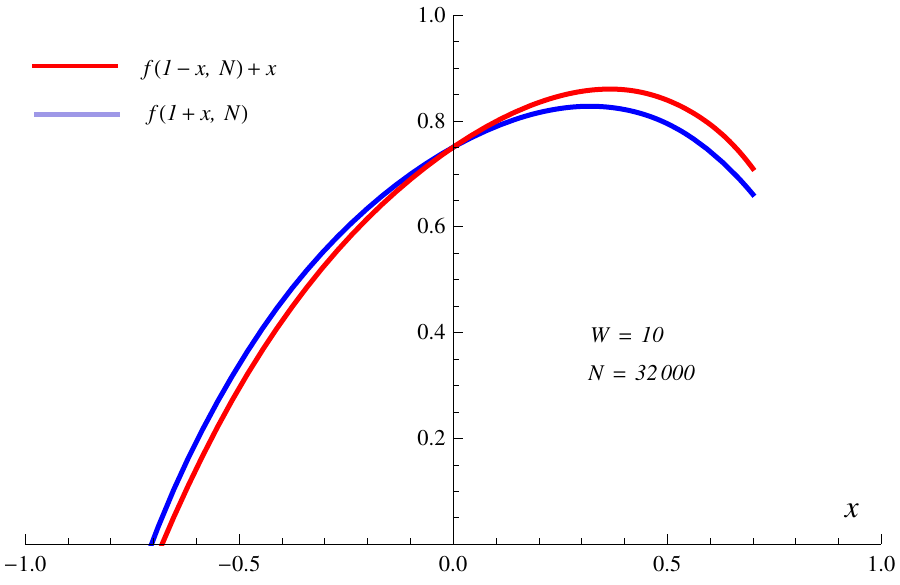}}
\caption{(Color online)
 Verification of the symmetry Eq.(\ref{sym-f-alpha}) for $f(\alpha,N)$ with $N=32000$ at disorder strength $W=10$.} \label{Fig:MF10}
\end{figure}

As a quantitative argument we show how the non-trivial symmetry Eq.(\ref{sym-f-alpha}) is fulfilled for $f(\alpha,N)$ in the region of extended states despite $\ln N\sim10$ is clearly not sufficient to identify $f(\alpha,N)=f(\alpha)$. In Fig.\ref{Fig:MF10} we plot $f(1+x,N)$ together with $f(1-x,N)+x$ for $W=10$ and $N=32000$. Albeit not perfect, the coincidence of the curves cannot be accidental.

Finally, Fig.~\ref{Fig:f-N-dep} shows the dependence of $f(\alpha,N)$ on the size of the system $N$ for disorder strength $W=10$ in the region of extended states. Remarkably, in all set of plots there are two apparent fixed points which positions depend on the disorder strength $W$.  If such apparent fixed points are the genuine fixed points then it is impossible for the $f(\alpha,N)$ curve to approach the GOE limiting curve concentrated near $\alpha=1$. This is the most solid numerical argument in favor of our conjecture that all extended states on the Bethe lattice are non-ergodic.

\section{Conclusion}
In this paper we introduced a new statistical measure of random wave functions --the support set-- and applied it to the study of the wave functions of the Anderson model on the Bethe lattice. As a guiding idea we exploited the multifractal ansatz Eq.(\ref{mult-anz}) for the wave function amplitude distribution function. We  derived certain relationships for exponents that describe scaling with the system size $N$ using the symmetry Eq.(\ref{MF-sym}) and checked them numerically. The fulfillment of these relationships and especially the presence of a fixed point on the dependence $f(\alpha,N)$ allows us to conjecture that in the entire delocalized region the extended states are non-ergodic, evolving from almost localized at the Anderson transition to almost ergodic at weak disorder.

{\it Acknowledgments.} A.S. would like to thank Y.V.\ Fyodorov for discussions at the beginning of this project, and the authors are grateful to Eugene Bogomolny and  Markus Mueller for useful discussions and  in particular to Giulio Biroli for discussions on his work \cite{Biroli} which was one of the main motivations for us to undertake the work discussed in this paper.
\bibliographystyle{apsrev}
\bibliography{../albethe.bib}

\end{document}